\documentclass[aps,prl,twocolumn,superscriptaddress,footinbib,notitlepage,floatfix]{revtex4-2}
\usepackage[latin9]{inputenc}
\usepackage{color}
\usepackage{amsmath}
\usepackage{multirow}
\usepackage{bbm}
\usepackage{amssymb}
\usepackage{makecell}
\usepackage{float}
\usepackage{booktabs}
\usepackage{pifont}
\usepackage{graphicx}
\usepackage[unicode=true,bookmarks=true,bookmarksnumbered=false,bookmarksopen=false,breaklinks=false,pdfborder={0 0 1},backref=false,colorlinks=true]{hyperref}
\hypersetup{colorlinks=true, linkcolor=blue, urlcolor=blue, citecolor=blue, pdfstartview={FitH}, hyperfootnotes=false, unicode=true}
\makeatletter
\renewcommand{\arraystretch}{1.35}
\usepackage[table,xcdraw]{xcolor}
\usepackage{framed}
\usepackage{fbox}
\usepackage{amsfonts}
\usepackage{tabularx}
\usepackage{dcolumn}
\usepackage{bm}
\usepackage{epstopdf}
\usepackage{times}
\hypersetup{urlcolor=blue}

\usepackage{tikz}
\usetikzlibrary{quantikz}
\usepackage{algorithm}
\usepackage{algorithmic}
\usepackage{amsthm}
\usepackage{booktabs,siunitx}

\definecolor{blue}{rgb}{0,0,1}
\definecolor{red}{rgb}{1,0,0}
\definecolor{green}{rgb}{0,1,0}

\renewcommand{\tablename}{Tab.}

\begin{document}

\title{Rate-1/2 Quantum LDPC Codes for Multi-tier Superconducting Hardware}

\author{Jiaxuan Zhang}
\email{zhangjx1@iai.ustc.edu.cn}
\affiliation{Institute of Artificial Intelligence, Hefei Comprehensive National Science Center, Hefei, Anhui, 230088, China}

\author{Zhao-Yun Chen}
\email{chenzhaoyun@iai.ustc.edu.cn}
\affiliation{Institute of Artificial Intelligence, Hefei Comprehensive National Science Center, Hefei, Anhui, 230088, China}

\author{Jia-Ning Li}
\affiliation{Laboratory of Quantum Information, University of Science and Technology of China, Hefei, 230026, China}
\affiliation{Anhui Province Key Laboratory of Quantum Network, University of Science and Technology of China, Hefei, 230026, China}

\author{Tian-Hao Wei}
\affiliation{Laboratory of Quantum Information, University of Science and Technology of China, Hefei, 230026, China}
\affiliation{Anhui Province Key Laboratory of Quantum Network, University of Science and Technology of China, Hefei, 230026, China}

\author{Qing-Yang Hou}
\affiliation{Laboratory of Quantum Information, University of Science and Technology of China, Hefei, 230026, China}
\affiliation{Anhui Province Key Laboratory of Quantum Network, University of Science and Technology of China, Hefei, 230026, China}

\author{Wei-Cheng Kong}
\affiliation{Origin Quantum Computing,  Hefei, China}

\author{Yu-Chun Wu}
\email{wuyuchun@ustc.edu.cn}
\affiliation{Laboratory of Quantum Information, University of Science and Technology of China, Hefei, 230026, China}
\affiliation{Anhui Province Key Laboratory of Quantum Network, University of Science and Technology of China, Hefei, 230026, China}
\affiliation{Institute of Artificial Intelligence, Hefei Comprehensive National Science Center, Hefei, Anhui, 230088, China}

\author{Guo-Ping Guo}
%\email{gpguo@ustc.edu.cn}
\affiliation{Laboratory of Quantum Information, University of Science and Technology of China, Hefei, 230026, China}
\affiliation{Anhui Province Key Laboratory of Quantum Network, University of Science and Technology of China, Hefei, 230026, China}
\affiliation{Institute of Artificial Intelligence, Hefei Comprehensive National Science Center, Hefei, Anhui, 230088, China}
\affiliation{Origin Quantum Computing,  Hefei, China}

\date{\today}

\begin{abstract}
Quantum error correction is essential for scalable quantum computing, while its resource overhead is governed in large part by the encoding rate, namely the number of logical qubits encoded per physical data qubit. Recent quantum low-density parity-check (qLDPC) constructions have entered the ultra-high-rate regime with encoding rates above 1/2, but their syndrome-extraction circuits generally rely on reconfigurable connectivity or long-range qubit transport, making them difficult to implement on fixed-connectivity superconducting hardware. Here we introduce Armillary codes, a family of ultra-high-rate qLDPC codes whose algebraic structure is generated by native iSWAP routing under hardware constraints. The construction jointly enforces stabilizer commutation and correct parallel syndrome extraction, while admitting radial and circumferential couplings in crossing-free routing layers. We construct explicit codes with rates above 1/2, stabilizer weight 12, Tanner-graph girth 6, and distances from 12 to 16, with layouts on 5 or 6 hardware tiers. Compared with the Gross and Mitten codes at the same distances, the Armillary codes achieve encoding rates about 6 and 2.5 times those of the Gross and Mitten codes, respectively, while maintaining lower hardware-complexity scores. In circuit-level simulations, the \([\![8688,4348,16]\!]\) code approaches the teraquop regime, achieving a logical error rate of \(1.62\times10^{-12}\) per logical qubit per round at a physical error rate of 0.1\%. These results show that the rate-\(1/2\) regime of qLDPC codes can be extended from reconfigurable platforms to superconducting hardware, providing a route toward low-overhead fault-tolerant quantum processors.
\end{abstract}

\noindent 

\maketitle

Quantum error correction (QEC) makes fault-tolerant quantum computing possible by encoding logical information redundantly across physical qubits~\cite{shor1995scheme}. For the past decade, the surface code has been the benchmark for QEC owing to its planar, nearest-neighbor layout and high threshold~\cite{dennis2002topological,fowler2012surface,google2025below}. Attention is now shifting toward quantum low-density parity-check (qLDPC) codes, which exploit nonlocal connectivity to achieve much lower overhead~\cite{bravyi2010tradeoffs,breuckmann2021quantum,panteleev2021degenerate}. The key driver of this shift is the encoding rate, that is, the number of logical qubits encoded per data qubit. The most recent frontier is the ultra-high-rate regime~\cite{kasai2026breaking,zhao2026ultrahighrate,yang2026gala,lee2026designprinciples}. Following Kasai's construction based on affine block matrices, several qLDPC codes have reached encoding rates above $1/2$~\cite{kasai2026breaking,zhao2026ultrahighrate}. These constructions, however, exploit highly flexible connectivity. The stabilizer checks require qubit permutations across the block, which are most naturally implemented by moving qubits. Consequently, rate-$1/2$ qLDPC codes have so far been proposed and benchmarked primarily for reconfigurable neutral-atom arrays, where atom transport supplies the requisite long-range connectivity~\cite{zhao2026ultrahighrate,yang2026gala,xu2024constantoverhead}.

Superconducting hardware is another leading QEC platform, with fast gates, high-fidelity readout, and a mature fabrication system~\cite{blais2021circuit,marxer2026fidelity,spring2025fast}. Yet the codes compatible with their fixed connectivity have remained at relatively modest rates. The Gross code~\cite{bravyi2024highthreshold} and Mitten codes~\cite{bhardwaj2026mitten} are representative examples that appeared successively, with data-qubit encoding rates of $1/12$ and $1/5$, respectively, substantially below the rate-$1/2$ neutral-atom results. The implementation of these codes is typically considered in multi-tier superconducting architectures. Experimentally, multi-tier architectures with superconducting bump bonds \cite{rosenberg20173d,field2024modular,kosen2024signal,norris2025performance} and through-silicon vias (TSVs) \cite{yost2020solid,mallek2021fabrication,hazard2023characterization} have been separately demonstrated. These capabilities make nonlocal qLDPC codes conceivable on superconducting hardware. 

The central challenge is to retain sufficient algebraic freedom for rate-1/2 encoding with useful distance while restricting the physical interactions to a layout with a small tier count, no crossings within a routing layer, and bounded coupling lengths. Table~\ref{tab:design-comparison} summarizes representative qLDPC codes against these three requirements. None of the prior proposals compared in Table~\ref{tab:design-comparison} has demonstrated all three requirements simultaneously. Kasai codes reach encoding rates above \(1/2\) but rely on reconfigurable connectivity, whereas the Gross and Mitten codes are compatible with superconducting platforms but remain at rates below \(1/2\).

\begin{table*}[t]
\centering
\hypersetup{urlcolor=black}
\small
\setlength{\tabcolsep}{5pt}
\renewcommand{\arraystretch}{1.15}
\begin{tabular*}{\textwidth}{@{\extracolsep{\fill}}llccc@{}}
\toprule
Code family & Target platform & \makecell{Encoding rate\\$R>1/2$} & \makecell{Low tier-count \\budget} & \makecell{Bounded length \\ budget}\\
\midrule
{Kasai codes} \cite{kasai2026breaking,zhao2026ultrahighrate} & Neutral atoms & {\color{black}\ding{51}} & {\color{black}\ding{55}} & {\color{black}\ding{55}} \\
{Gross code} \cite{bravyi2024highthreshold} & Superconducting & {\color{black}\ding{55}} & {\color{black}\ding{51}} & {\color{black}\ding{51}} \\
{Mitten codes} \cite{bhardwaj2026mitten} & Neutral atoms / superconducting & {\color{black}\ding{55}} & {\color{black}\ding{51}} & {\color{black}\ding{51}} \\
\midrule
\textbf{Armillary codes} (this work) & Superconducting & {\color{black}\ding{51}} & {\color{black}\ding{51}} & {\color{black}\ding{51}} \\
\bottomrule
\end{tabular*}
\caption{\textbf{Comparison of Representative qLDPC codes.} Several existing qLDPC codes are compared with the Armillary codes under the three design requirements. For the two hardware-layout criteria, a cross indicates that an explicit layout meeting the criterion has not been demonstrated in the cited proposal.}
\label{tab:design-comparison}
\end{table*}

A promising way has emerged from the native gate set of superconducting hardware~\cite{mcewen2023timedynamics,geher2025directional}. Rather than realizing a QEC code on a static connectivity graph, iSWAP-based schemes define the code dynamically, where an ancilla is routed through the data array, and the stabilizer support is generated by the path \cite{geher2025directional,zhang2026routing}. Directional codes, routing codes, and related tile constructions have exploited this dynamic connectivity to achieve higher encoding rates than static codes under the same hardware constraints~\cite{geher2025directional,zhang2026routing,gu2026nearest,nixon2026vine}. These results suggest that the native iSWAP interaction can be elevated from a routing primitive to a code-design resource, providing a way to circumvent the connectivity bottleneck of solid-state platforms.

In this work, based on iSWAP gates and through a distinct algebraic construction, we introduce \textit{Armillary codes}, a novel family of qLDPC codes that achieves encoding rates above 1/2 and is suitable for multilayer architectures. Group-dependent channel permutations and bounded shifts generate the blocks of non-orthogonal parent matrices, while constraints on the retained checks jointly ensure CSS commutation and correct parallel syndrome extraction. This construction achieves encoding rates above 1/2 with explicit multi-tier layouts. Algebraically, the key difference between our codes and Kasai codes is that we do not require preassigned affine permutation blocks. Our construction uses non-orthogonal parent matrices without requiring a prescribed affine or block-circulant form. In our construction, the native permutation of the iSWAP (CXSWAP) gate generates the blocks required by parent matrices, and these permutations are fixed by the iSWAP operations that the hardware can execute natively rather than by an algebraic pre-assignment. 

Concretely, this construction yields explicit codes with rates above $1/2$ and distances from 12 to 16, including $[\![3636,1822,12]\!]$, $[\![4932,2470,14]\!]$, and $[\![8688,4348,16]\!]$. All three have check weight twelve, Tanner-graph girth six, and complete a full syndrome-extraction cycle within 12 layers of two-qubit gates. On the hardware-layout side, we assume a finite number of long-range couplers routed within a layer, no crossings within a layer, and a coupling length much shorter than the chip's global dimension. These assumptions are consistent with the constraints of multi-tier superconducting hardware. Remarkably, under these assumptions, the connectivity required by Armillary codes fits on only 5 or 6 hardware tiers. At the same distances, the Armillary codes achieve lower hardware-complexity scores under hardware-aware-layout (HAL)~\cite{halpaper} metrics while raising the encoding rate to 2.5 times that of the Mitten codes. For the distance-12 and distance-14 instances, the hardware-complexity scores are even lower than that of the $[\![144,12,12]\!]$ Gross code. In circuit-level simulations, the $[\![8688,4348,16]\!]$ code achieves a logical error rate of $1.62\times 10^{-12}$ per logical qubit per round at physical error rate of 0.1\%, approaching the teraquop regime. To our knowledge, this is the first rate-$1/2$ qLDPC codes with an explicit multi-tier superconducting layout to reach such logical error rates.

Although experimental implementation challenges remain, we believe these results provide a practical path to ultra-high-rate QEC on fixed-connectivity superconducting hardware. The rate-$1/2$ regime, previously accessible only to reconfigurable neutral-atom platforms, becomes in principle accessible to multi-tier superconducting processors, which keeps superconducting architectures competitive in the fault-tolerant quantum computing.

\vspace{.4cm}
\noindent\textbf{\large{}Code construction} \\
\noindent  

A $[\![n,k,d]\!]$ quantum code encodes $k$ logical qubits into $n$ data qubits, and its distance $d$ is the minimum weight of a nontrivial logical operator. The encoding rate is defined as $R=k/n$. Here we focus on CSS codes~\cite{calderbank1996good,steane1996error}, whose stabilizer generators are purely $X$ type or $Z$ type. The rows of the check matrices $H_X$ and $H_Z$ correspond to these generators and satisfy $H_XH_Z^{\mathsf T}=0$. The code space is the simultaneous $+1$ eigenspace of the stabilizer generators. To measure a stabilizer generator, one usually prepares a syndrome qubit in the appropriate basis, applies the CX gate between the syndrome qubit and the data qubits in its support, and measures the syndrome qubit.

Our stabilizer measurement circuit is based on the dynamic circuit with iSWAP gates, which are locally equivalent to CXSWAP gates. A CXSWAP applies a CX and then swaps the states of the two positions, ${\mathrm{CXSWAP}}=\mathrm{SWAP}\circ\mathrm{CX}$. The syndrome qubit therefore moves through the data array, and the support of the measured stabilizer is generated by the path that the syndrome qubit traverses. Here, data qubits and syndrome qubits are labeled by their functional roles. A position refers to the physical qubit that carries a given role at a given time step. The same position can carry different roles at different times. These dynamic circuits have been comprehensively described in Refs.~\cite{geher2025directional,zhang2026routing,gu2026nearest}. Our focus on this framework is motivated by its ability to simplify the hardware connectivity, reducing both the connectivity degree and the connection length relative to a static stabilizer measurement circuit~\cite{mcewen2023timedynamics,zhang2026routing,gu2026nearest,zhou2026louvre}.

Increasing the encoding rate requires reducing the number of independent stabilizer constraints. Our construction starts from non-orthogonal parent matrices $\hat{H}_X$ and $\hat{H}_Z$, whose rows are candidate checks, and obtains check matrices $H_X$ and $H_Z$ with a sufficiently large rate by deleting block rows. We use Kasai's principle of imposing orthogonality only on the retained checks~\cite{kasai2026breaking}. Here, those checks must emerge from hardware-permitted CXSWAP paths that also satisfy the ordered-interaction conditions for parallel measurement.

\begin{figure*}[t]
\centering
\includegraphics[width=\textwidth]{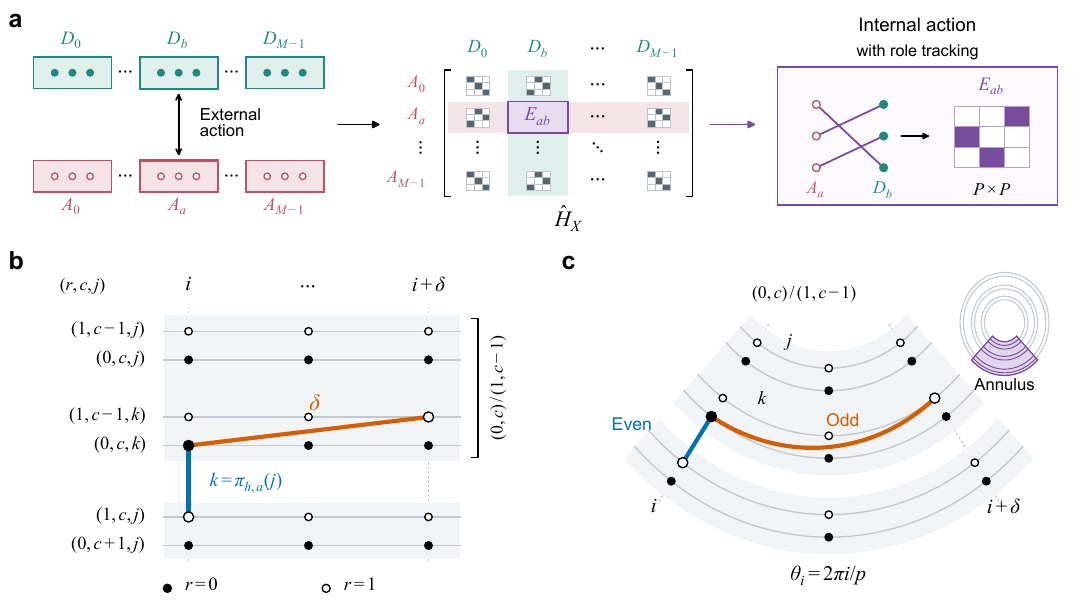}
\caption{\textbf{Code construction framework and geometric layout.}
\textbf{a,} Data and syndrome qubits are partitioned into groups $D_b$ and $A_a$ of $P$ qubits (left). The same groups label the block columns and rows of the parent matrix (middle), and the highlighted pair selects $E_{ab}$. The external action pairs groups, and the internal action pairs physical positions within the matched groups. Tracking the transported roles yields the effective matching $e_{ab}=\beta_{b,t}^{-1}\circ f_{a,t}\circ\alpha_{a,t}$ shown at right. Miniature matrices represent permutation blocks schematically.
\textbf{b,} Selected physical positions in an unfolded strip. Rail labels $(r,c,j)$ omit the station coordinate $i$, which varies along each rail. Each position is fully specified by $(r,c,i,j)$. Filled and open dots denote $r=0$ and $r=1$. The upper shaded band interleaves $(0,c,j)$ with $(1,c-1,j)$, and the lower pair continues the same pattern with $(0,c+1,j)$ and $(1,c,j)$. Thus, the two rails of a physical module carry different values of $c$.
\textbf{c,} The same positions and pairings are shown on an annulus, with the displayed sector marked in the overview. The even step connects $(1,c,i,j)$ to $(0,c,i,k)$, where $k=\pi_{h,a}(j)$ (blue). The following odd step reaches $(1,c-1,i+\delta,k)$ (orange).}
\label{fig:construction}
\end{figure*}

To realize a block structure in such circuits, we fix both the row groups and the column groups of the parent matrix. The row groups come from a grouping of the syndrome qubits, and the column groups come from a grouping of the data qubits. Both groupings are defined by the qubit roles, not by fixed physical positions. The data and syndrome qubits are each partitioned into $M$ groups, $D_0,\ldots,D_{M-1}$ and $A_0,\ldots,A_{M-1}$, of size $P$, so $n=MP$. At each step of the stabilizer measurement circuit, a bijection is established between the $M$ syndrome groups and the $M$ data groups. We call this bijection the external action. Within each pair of groups, a bijection between the $P$ syndrome qubits and the $P$ data qubits determines which pairs of qubits the CXSWAP acts on. We call this bijection the internal action. The correspondence between these groupings, the two actions, and the blocks of the parent matrix is illustrated in \hyperref[fig:construction]{Fig.~\ref*{fig:construction}(a)}.

At step $t$, suppose that syndrome group $a$, of $X$ type, performs an external action with data group $b$. Let $\alpha_{a,t}$ and $\beta_{b,t}$ denote the bijections from the qubits of the two groups to their positions at that time, and let $f_{a,t}$ denote the internal action. The data qubit that syndrome qubit $u$ actually acts on is 
\begin{equation} 
e_{ab} (u)=\beta_{b,t}^{-1}\left(f_{a,t}\left(\alpha_{a,t}(u)\right)\right). 
\label{eq:effective}
\end{equation}
Here, \(\alpha_{a,t}\) and \(\beta_{b,t}\) track the cumulative role permutations before step \(t\), whereas \(f_{a,t}\) specifies the physical pairing at that step.

Then we define a $P\times P$ permutation matrix $E_{ab}$ by $(E_{ab})_{u,v}=1$ if
and only if $v=e_{ab}(u)$. If each syndrome group visits each initial data
group exactly once, the parent matrix takes the block form
\begin{equation}
\hat{H}_X=
\begin{pmatrix}
E_{00} & E_{01} & \cdots & E_{0,M-1}\\
E_{10} & E_{11} & \cdots & E_{1,M-1}\\
\vdots & \vdots & \ddots & \vdots\\
E_{M-1,0} & E_{M-1,1} & \cdots & E_{M-1,M-1}
\end{pmatrix}.
\label{eq:parent}
\end{equation}
This extra assumption ensures that each block is a single permutation matrix $E_{ab}$, rather than a sum of permutation matrices. The $Z$-type parent matrix has the same block form. The difference from Kasai codes is now clear. The permutation blocks need not be preassigned affine permutations. They are determined entirely by the external and internal actions, without requiring an overall block-circulant form.

After constructing the parent matrix, we verify the orthogonality of the retained checks and delete block rows accordingly. The required orthogonality has two aspects. We write the check action at each step as $H_X(t)$ and $H_Z(t)$, so that $H_X=\sum_t H_X(t)$ and $H_Z=\sum_t H_Z(t)$, and define
\begin{equation}
\Omega=H_X H_Z^{\mathsf T},\qquad
C=\sum_{0\le t<u<M}H_X(t)H_Z(u)^{\mathsf T}.
\label{eq:two}
\end{equation}
The condition $\Omega=0$ ensures that the check matrices commute, while $C=0$ further ensures that the stabilizers are measured correctly in the parallel stabilizer measurement circuit. 

Deleting a block row means that the corresponding group of candidate checks is not measured. This can be achieved by setting the corresponding syndrome qubits to $|0\rangle$ and then applying CXSWAP to realize an equivalent SWAP operation. This means that the syndrome qubits at these positions are used exclusively for routing. The remaining checks must commute. Reducing the number of independent retained constraints increases the number of logical qubits, but preserving a target distance requires excluding all lower-weight operators that commute with the retained checks and lie outside their stabilizer group.

So far, the framework still leaves a large space of admissible codes and therefore calls for a concrete construction tailored to the constraints of geometric layout. We fix the number of groups $M$, which is also the number of time steps and the weight of every stabilizer, since each syndrome group visits each data group exactly once. Each position is labeled by $(r,c,i,j)$, where $r\in\{0,1\}$ records the initial role ($r=0$ for data, $r=1$ for syndrome), $c\in\mathbb{Z}_M$ is the group index, and $i \in \mathbb{Z}_p$, $0\le j<q$ label the qubit within the group. Each group therefore contains $P=qp$ positions. We refer to the $p$ values of $i$ and $q$ values of $j$ as stations and channels, respectively. The four coordinates and their relation are shown in \hyperref[fig:construction]{Fig.~\ref*{fig:construction}(b)}.

Then, we adopt the following two key rules for the construction. At even step $t=2h$, the syndrome qubits of group $a$ are at $r=1$, and the CXSWAP pairing is
\begin{equation}
(1,c,i,j)\longleftrightarrow(0,c,i,\pi_{h,a}(j)).
\label{eq:even}
\end{equation}
At odd step $t=2h+1$, the syndrome qubits of group $a$ are at $r=0$, and the CXSWAP pairing is
\begin{equation}
(0,c,i,j)\longleftrightarrow(1,c-1,i+\delta_{h,a,j},j).
\label{eq:odd}
\end{equation}
The map $\pi_{h,a}$ permutes the $q$ channels within a group at fixed $c$ and $i$. It acts at even step $2h$ on syndrome group $a$, and it does not depend on $i$, so all $p$ positions share the same permutation. The integer $\delta_{h,a,j}$ is a signed shift of the index $i$, applied at odd step $2h+1$ to syndrome group $a$ and the current channel $j$. These two rules define the bijections $\alpha_{a,t}$, $\beta_{b,t}$, and $f_{a,t}$ of the previous section.

Let us explain the rationale behind the rules in terms of the external and internal actions, respectively. The restrictions on $c$ determine the external action. Suppose that a data qubit initially at $(0,-b,i,j)$ and a syndrome qubit initially at $(1,-a,i',j')$ interact at step $t$. Because the syndrome group index decreases by one every two steps while the data group index increases by one, we have
\begin{equation}
b = a+t\pmod M.
\label{eq:visit}
\end{equation}
In other words, each syndrome group $a$ visits each data group $b$ exactly once, as required.

The internal action is restricted to a permutation $\pi_{h,a}$ of the $q$ channels combined with a translation $\delta_{h,a,j}$ of the index $i$. These maps form the semidirect product $S_q\ltimes C_p^q$, where \(C_p\) is the cyclic group of order \(p\), generated by the channel permutations and the translations. This is much smaller than the full symmetric group on $P$ positions, which makes the search tractable.

The geometry of this construction is naturally drawn on an annular layout, as shown in \hyperref[fig:construction]{Fig.~\ref*{fig:construction}(c)}. A folded ordering determines the radial positions associated with the index $c$, the $p$ values of $i$ are angular positions $\theta_i=2\pi i/p$, and the $q$ channels are concentric tracks. In particular, the physical module containing the $r=0$ rail of column $c$ also contains the $r=1$ rail of column $c-1$. These rails are interleaved across the channels. The even-step rule connects the positions $(1,c,i,j)$ and $(0,c,i,\pi(j))$. This connection is radial, with no angular displacement. The odd-step rule connects the positions $(0,c,i,j)$ and $(1,c-1,i+\delta,j)$. This connection is circumferential, with angular displacement $|\delta|\beta$ and radial displacement $w$, where $\beta=2\pi/p$ and \(w\) is the local radial spacing, which can vary across modules and channels in subsequent layout optimization. Its displacement in the unfolded angular coordinate is $|\delta|s$, with $s=R_0\beta$ the arc pitch at the reference radius $R_0$. The two families are nearly orthogonal in the plane of radius and angle, providing simple local paths that can be distributed across noncrossing routing layers in the multi-tier structure.

Finally, orthogonality is enforced through a block-deletion principle. For a chosen set of retained \(X\)- and \(Z\)-type block rows, we pair the column blocks so that paired blocks carry identical relative actions and lie on the same side of the cyclic ordering, which guarantees cancellation in both \(\Omega\) and \(C\) over \(\mathbb{F}_2\) and hence \(\Omega=C=0\). The pairing condition separates into a channel constraint on the channel permutations and an angular constraint on the accumulated shifts. Once the channel permutations are fixed, the angular constraint reduces to a linear system over the shifts. We retain only those block-deletion configurations for which both constraints admit solutions, so that the remaining checks satisfy \(\Omega=C=0\) and constitute a valid parity-check matrix. Full details are provided in the Methods. We then evaluate the encoding rate from the ranks of the retained check matrices and search for low-weight nontrivial logical operators. Instances with rates above $1/2$ that meet the target distance are retained as candidates for further benchmarking.

\vspace{.4cm}
\noindent\textbf{\large{}Explicit instances and logical performance} \\

We searched over the cases \(M=12\) with \(q=3\) and \(q=4\), and obtained a family of codes with rates above \(1/2\). Three representative codes are \([\![3636,1822,12]\!]\), \([\![4932,2470,14]\!]\), \([\![8688,4348,16]\!]\). The two \(q=3\) codes use \(p=101\) and \(p=137\), with \(P=303\) and \(P=411\) and the \(q=4\) code uses \(p=181\) and \(P=724\). We further verified that the X- and Z-check Tanner graphs of each code have girth 6. Additional codes from the search and reproducibility details for these codes are given in the Supplementary Information~\cite{SupplementaryInformation} and \cite{ArmillaryCodeRepository}.

We benchmarked the performance of the three codes under circuit-level noise. Each two-round circuit consists of a forward round and a reverse round, with the latter using the reversed CXSWAP sequence, following the standard scheme for dynamical QEC codes~\cite{geher2025directional,zhang2026routing,shaw2026morphing}. We assume that the elementary operations of the circuit are CXSWAP gates together with preparation and measurement in the $X$ and $Z$ bases. Single-qubit depolarizing uniform channels are inserted after single-qubit operations and idle steps, and before measurement, while two-qubit depolarizing uniform channels are inserted after two-qubit gates, all with noise strength $p_{\mathrm{phy}}$.

We employ a two-stage decoding algorithm that combines relay-BP~\cite{muller2025relay} with a most-likely-error (MLE) decoder~\cite{landahl2011color}. In the relay-BP stage, we first run a short-budget relay-BP decoder. If its output is consistent with the observed syndrome, we accept it. Otherwise, the decoder is rerun with a long budget starting from the original prior. If neither run yields a syndrome-consistent output, we proceed to the second stage. The second stage solves a mixed-integer program that finds the most likely error consistent with the syndrome. The parameters of the relay-BP and the details of the MLE decoder are presented in the Methods.

\begin{figure}[t]
\centering
\includegraphics[width=\columnwidth]{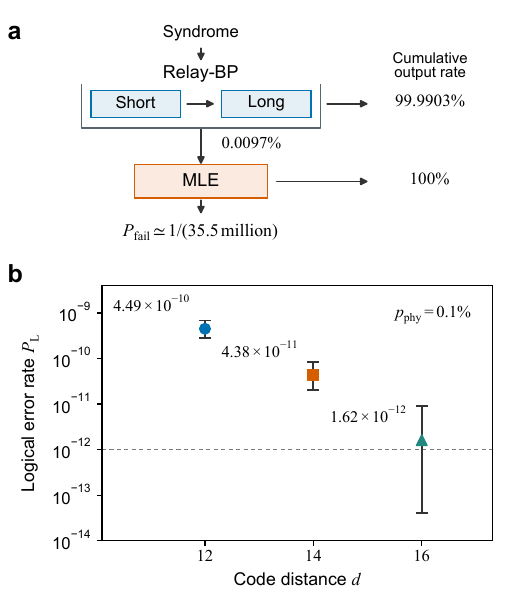}
\caption{\textbf{Decoding and circuit-level performance.}
\textbf{a,} Two-stage decoding with relay-BP and MLE. The proportions shown are for the $[\![8688,4348,16]\!]$ code. The downward percentage gives the fraction passed to MLE, and right-hand percentages give cumulative syndrome-consistent output rates. $P_{\mathrm{fail}}$ is the probability that at least one logical error remains in a code block after decoding, including errors from either stage.
\textbf{b,} Symbols give the logical error rate $P_{\mathrm L}$ per logical qubit per QEC round at $p_{\mathrm{phy}}=0.1\%$, with equally sampled $X$ and $Z$ bases. Whiskers give nominal 95\% binomial reference intervals. The dashed line marks $10^{-12}$. Sample counts and statistical details are given in the Supplementary Information~\cite{SupplementaryInformation}.}
\label{fig:decoding}
\end{figure}

At $p_{\mathrm{phy}}=0.1\%$, relay-BP returns a syndrome-consistent output for more than $99.99\%$ of samples, leaving fewer than $10^{-4}$ to MLE, as illustrated in \hyperref[fig:decoding]{Fig.~\ref*{fig:decoding}(a)}. Incorrect syndrome-consistent outputs are included in the final error rate even though they do not trigger MLE. \hyperref[fig:decoding]{Fig.~\ref*{fig:decoding}(b)} shows the logical error rates per logical qubit per QEC round, normalized according to Eq.~(\ref{eq:rates}): $4.49\times10^{-10}$, $4.38\times10^{-11}$, and $1.62\times10^{-12}$ for distances 12, 14, and 16, respectively. For the distance-16 code, the residual block failure probability is approximately one in 35.5 million. Across the three tested codes, all failures originate in relay-BP, indicating that future improvements to the early-stage decoder may further reduce the logical error rate. Its point estimate approaches the teraquop regime, with counting uncertainty shown in the figure.

\vspace{.4cm}
\noindent\textbf{\large{}Implementation on multi-tier superconducting hardware} \\
\noindent 
We consider a multi-tier superconducting architecture under assumptions consistent with Ref.~\cite{halpaper}. The qubit tier at the bottom corresponds to a single routing layer, while each higher tier provides both its top and bottom surfaces as wire layers, with bump bonds connecting opposing planes within a tier, TSVs connecting different tiers, and no wire crossings within any layer.

The proposed code is compatible with these constraints because its connections are approximately radial and circumferential on a ring. The even-step rule produces radial edges, and the odd-step rule produces circumferential edges. This structure provides local routing paths whose shapes can be adjusted without changing the circuit. We retain the annular ordering of the qubits and optimize how these paths share routing faces and how the faces are stacked (see Methods).

Table~\ref{tab:hal} compares our layouts with those of the Gross and Mitten codes, reporting the number of tiers, the Length, TSVs, and Bonds metrics, and the hardware-complexity score. We follow HAL's averaging rules, using actual arc lengths normalized by the minimum qubit pitch. Our length measure accounts for the full geometric length of diagonal segments, making it stricter than the equal-step counting used in the published Gross and Mitten code results. The distance-12 and distance-14 layouts use five tiers, while the distance-16 layout uses six. Their hardware-complexity scores are 1.78, 1.81, and 2.17, respectively, all below those of the Mitten layouts at the same distances in the table, with encoding rates about 2.5 times as high. The distance-12 and distance-14 scores are also below that of the \([\![144,12,12]\!]\) Gross code, while the encoding rate is about six times as high. Their low Bonds values reflect the absence of intermediate face switches along the in-plane paths. 

\begin{table}[t]
  \centering
  \small
  \setlength{\tabcolsep}{6pt}
  \renewcommand{\arraystretch}{1.08}
  \begin{tabular}{@{}l c c c c c@{}}
    \toprule
    Code & Tiers & Length & TSVs & Bonds & \(C_{\mathrm{hw}}\) \\
    \midrule
    \multicolumn{6}{@{}l}{\textbf{Armillary codes}}\\[1pt]
    \([\![3636,1822,12]\!]\) & 5  & \phantom{0}7.61 & 3.42 & 0.91 & 1.78 \\
    \([\![4932,2470,14]\!]\) & 5  & \phantom{0}8.59 & 3.43 & 1.01 & 1.81 \\
    \([\![8688,4348,16]\!]\) & 6  & 15.75 & 4.43 & 1.23 & 2.17 \\
    \addlinespace[4pt]
    \multicolumn{6}{@{}l}{\textbf{Gross codes}}\\[1pt]
    \([\![144,12,12]\!]\)   & 5  & 11.08 & 3.27 & 5.06 & 2.12 \\
    \([\![288,12,18]\!]\)   & 5  & 13.94 & 3.75 & 5.13 & 2.24 \\
    \addlinespace[4pt]
    \multicolumn{6}{@{}l}{\textbf{Mitten codes}}\\[1pt]
    \([\![200,40,12]\!]\)   & 6  & \phantom{0}8.38  & 3.77 & 4.12 & 2.09 \\
    \([\![300,60,14]\!]\)   & 7  & \phantom{0}9.40  & 4.57 & 4.69 & 2.28 \\
    \([\![500,100,16]\!]\)  & 9  & 12.17 & 5.87 & 4.75 & 2.60 \\
    \bottomrule
  \end{tabular}
\caption{\textbf{Hardware complexity comparison using HAL metrics.} For our codes, length is the tier-averaged arc length normalized by the minimum qubit pitch. Gross and Mitten codes retain their published Length and \(C_{\mathrm{hw}}\). HAL counts axial and diagonal grid steps equally, giving a less stringent length assessment and making our comparison conservative. TSVs is averaged over higher-tier connections. Bonds is the maximum tier-average face-switch segment count in the HAL implementation, rather than a physical bump-bond count. The data for the Gross and Mitten codes were obtained using the HAL program and are taken from Ref.~\cite{halpaper} and Ref.~\cite{bhardwaj2026mitten}, respectively.}
\label{tab:hal}
\end{table}

Experimental advances in flip-chip interconnects, superconducting TSVs, and long-range couplers provide component-level support for this architecture~\cite{rosenberg20173d,yost2020solid,xu2026tunable,xiong2026scalable}. Integration across five or six tiers requires control of interconnect loss, parasitic modes, crosstalk, and fabrication yield~\cite{huang2021microwave,kosen2024signal,norris2025performance}. Our repeated routing motifs support reuse of coupler designs, while fixed layer assignments eliminate intermediate transitions needed to resolve crossings. Our layout specifies the spacing between wires, while connections near each qubit, shielding against unwanted coupling, and control and readout wiring still require detailed design~\cite{huang2021microwave,kosen2024signal,spring2025fast}.

\vspace{.4cm}
\noindent\textbf{\large{}Discussion} \\
Armillary codes show that high-rate qLDPC memories can be co-designed with superconducting hardware constraints. The CXSWAP operations that transport syndrome qubits determine the permutation blocks in the check matrices and the physical interaction paths. This co-design yields explicit \(R>1/2\) memories with finite-tier layouts and good circuit-level performance.  In particular, the distance-16 instance approaches the teraquop regime in simulation, suggesting a practical route to lower-overhead fault-tolerant superconducting processors.

Our search covers only a restricted part of the construction space. We fix \(M=12\), consider \(q=3\) and \(q=4\), and use a prescribed cyclic schedule, bounded shifts, selected block deletions, and a fixed class of annular layouts. These choices make the search manageable, but they do not imply that the resulting codes are optimal. Larger values of \(M\) and \(q\), more general external and internal actions, other retained check sets, and a joint search over code parameters and hardware layouts may produce codes with larger distance, higher rate, better logical performance, or lower hardware cost.

In addition, logical operations remain open. Our construction provides a memory protocol and a syndrome-extraction circuit, but not yet logical measurements and gates that preserve the layout advantages. Code surgery~\cite{cowtan2024surgery,cowtan2026parallel,maskara2026efficientlogic} is in principle feasible, yet it requires ancilla regions~\cite{huang2023homomorphic,williamson2026gauging} tailored to the characteristics of the dynamic circuits. A key question is how much extra coupling resource code surgery requires. If the existing tier, crossing, and length budgets can be largely preserved, Armillary codes could bridge high-rate quantum memories and a complete fault-tolerant processor architecture.

\vspace{.2cm}\noindent
\noindent\textbf{\large{}Code availability}\\The source codes for Armillary code construction and numerical simulations are available at \cite{ArmillaryCodeRepository}.

\vspace{.2cm}\noindent
\noindent\textbf{\large{}Acknowledgments}\\
The authors thank the Cardinal Optimizer (COPT)~\cite{copt} for solving the mixed-integer programs in this work. This work was supported by the Young Scientists Fund (C) of the National Natural Science Foundation of China (Grant No. 12604755), the National Key Research and Development Program of China (Grant No. 2023YFB4502500), Anhui Province Innovation Plan for Science and Technology (Grant No. 202423r06050002), and the China Postdoctoral Science Foundation (Grant No. 2026M793680).

\bibliography{ref}

\renewcommand{\tablename}{Extended Tab.}
\setcounter{table}{0}

\clearpage
\noindent\textbf{\large{}Methods}\\
\noindent\textbf{Code construction details}\\
\vspace{.2cm}
The construction uses an even number $M$ of data and syndrome groups, each containing $P=qp$ states, with $n=Mqp$ data qubits and $2n$ physical positions. The external schedule fixes which groups interact, while the internal actions are specified by channel permutations $\pi_{h,a}$ and bounded shifts $\delta_{h,a,j}$, with $h=0,\ldots,M/2-1$ and $a\in\mathbb Z_M$. Tracking the transported roles determines the effective permutation blocks $E_{ab}$. The representative codes retain three block rows in $X$ or $Z$ syndrome, giving check matrices of size $3qp\times Mqp$ and a design rate $1-6/M$. We now give the pairing constraints that ensure both commutation and ordered cancellation, and describe the resulting finite parameter search.

We first describe the orthogonality condition and the block deletion principle. For the retained \(X\)-type and \(Z\)-type block rows \(X\) and \(Z\), we define, for each \(x\in X\) and \(z\in Z\),
\begin{equation}
\begin{aligned}
K_b^{xz}&=E_{xb}E_{zb}^{\mathsf T},\qquad \\
\chi^{xz}(b)&=\mathbf{1}\left\{(b-x)\bmod M<(b-z)\bmod M\right\},
\end{aligned}
\label{eq:Kchi}
\end{equation}
where \(E_{xb}\) (\(E_{zb}\)) is the block of the parent matrix \(\hat{H}_X\) (\(\hat{H}_Z\)) at block rows \(x\) (\(z\)) and block column \(b\), and \(\mathbf{1}\{\cdot\}\) is the indicator function, equal to one when the condition holds and zero otherwise.

For each row pair \((x,z)\), we pair the \(M\) column blocks \(b=0,\ldots,M-1\) into \(M/2\) pairs \((b,b')\) such that
\begin{equation}
K_b^{xz}=K_{b'}^{xz},\qquad
\chi^{xz}(b)=\chi^{xz}(b'),
\label{eq:pairing}
\end{equation}
Eq.~\eqref{eq:pairing} is a sufficient condition for $\Omega=C=0$, since each pair contributes $2K_b=0$ to $\Omega$ and either twice or not at all to $C$, so both sums vanish over $\mathbb{F}_2$. Note that different \((x,z)\) may use different column pairings.

The condition \(K_b^{xz}=K_{b'}^{xz}\) is equivalent to
\begin{equation}
\left(E_{xb}\right)^{-1}E_{xb'}=\left(E_{zb}\right)^{-1}E_{zb'}.
\label{eq:relative}
\end{equation}
Both sides of Eq.~\eqref{eq:relative} are determined by internal actions, so the equation can split into a channel part and an angular part. Write the effective block action as $e_{ab}(i,j)=(i+\ell_{abj},\sigma_{ab}(j))$. The channel part is
\begin{equation}
\tau_b^{xz}:=\sigma_{zb}^{-1}\circ\sigma_{xb}
=\sigma_{zb'}^{-1}\circ\sigma_{xb'}=:\tau_{b'}^{xz},
\label{eq:channel}
\end{equation}
where \(\sigma_{ab}\) is the channel permutation of block \(E_{ab}\), and \(\tau_b^{xz}\) maps an \(X\)-side input channel to the \(Z\)-side input channel reaching the same data channel. We solve the channel constraints using the \(q!\) possible channel permutations. With the channel permutations fixed, the angular part is linear in the shifts. Then let \(\boldsymbol{\delta}\) collect all shifts \(\delta_{h,a,j}\), and let \(\ell_{abj}(\boldsymbol{\delta})\) be the accumulated angular shift of block \(E_{ab}\) on input channel \(j\). Each \(\ell_{abj}(\boldsymbol{\delta})\) is linear in \(\boldsymbol{\delta}\) since the transported coordinates accumulate signed odd-step shifts. Equating the relative angular shifts after this channel relabeling gives, for all \(x\in X\), \(z\in Z\), pairs \((b,b')\), and channels \(j\),
\[
\begin{aligned}
&\ell_{xbj}(\boldsymbol{\delta})-\ell_{xb'j}(\boldsymbol{\delta})\\
&\quad-\ell_{zb,\tau_b^{xz}(j)}(\boldsymbol{\delta})
+\ell_{zb',\tau_b^{xz}(j)}(\boldsymbol{\delta})=0\pmod p.
\end{aligned}
\]
Since every \(\ell\) is linear in \(\boldsymbol{\delta}\), these conditions are linear in \(\boldsymbol{\delta}\). Collecting them gives $A\boldsymbol{\delta}=0\pmod p$, where \(A\) is fixed once the channel permutations and column pairing are chosen. We retain those block-deletion configurations under which Eq.~(\ref{eq:channel}) and this linear equation have solutions. Then the remaining checks satisfy \(\Omega=C=0\), thus constitute a valid parity-check matrix.

For $M=12$, the six even--odd step pairs and 12 syndrome groups require only 72 channel permutations. In the $q=3$ construction, the channels share one shift $\delta_{h,a,j}=\delta_{h,a}$ for each $(h,a)$, giving 72 integer variables in $[-4,4]$. In the $q=4$ construction, the shifts depend on the current channel, giving 288 integer variables in $[-6,7]$. These signed ranges are fixed by the routing library, and all shifts act modulo $p$. Thus, for fixed $p$, retained block rows, and column pairings, the search is parameterized by 72 permutations and 72 or 288 bounded integers, rather than by the entries of the full check matrices, which greatly reduces the search space relative to searching over full check matrices. Before imposing the pairing constraints, the corresponding parameter spaces contain $(3!)^{72}9^{72}$ and $(4!)^{72}14^{288}$ assignments, respectively. The channel constraints and $A\boldsymbol{\delta}=0\pmod p$ restrict these assignments further.

\vspace{.2cm}
\noindent\textbf{Noise model and simulations}\\
\vspace{.2cm}
Let \(p_{\mathrm{phy}}\) denote the physical noise strength. For \(a=1,2\) qubits we use the uniform depolarizing channel
\begin{equation}
D_a(\rho)=(1-p_{\mathrm{phy}})\rho+\frac{p_{\mathrm{phy}}}{4^a-1}
\sum_{P\in\{I,X,Y,Z\}^{\otimes a}\setminus\{I^{\otimes a}\}}P\rho P.
\label{eq:depol}
\end{equation}

At the start of each QEC round, all \(n\) ancilla transport roles are reset, including routing roles that carry no retained check, and an ideal Hadamard is applied to the \(X\)-check syndromes, with the noisy preparation channel placed between the reset and this basis rotation. In each of the 12 steps, the required CXSWAP is applied on the physical positions, followed by one \(D_2\) channel. Deleting a check removes only the CX while keeping the SWAP part and its gate noise. At the end of the round, an ideal basis rotation is applied to the \(X\)-syndrome qubits, one \(D_1\) channel is applied to the \(n/2\) check syndromes that are measured, followed by a \(Z\)-basis measurement, and the data \(D_1\) channel is placed at the round end to account for the idle between syndrome measurement and the next preparation. 

The memory experiment simulation consists of $n_r=4$ noisy QEC rounds, comprising two complete forward-reverse cycles, followed by one ideal round. In the rare-failure regime, the accumulated memory failure probability is approximately proportional to the number of QEC rounds when the logical failure process is approximately stationary. We use this fixed window to devote the available computational budget to sampling rare logical failures, and report the logical error rate per logical qubit per round using Eq.~(\ref{eq:rates}). The experiment is repeated for sampling, with half of the samples prepared and measured in the \(X\) basis and the other half in the \(Z\) basis. For each code and each memory basis, the detector error model is generated using Stim~\cite{gidney2021stim}. 

\vspace{.2cm}
\noindent\textbf{Decoding strategy}\\
\vspace{.2cm}

The decoder consists of two stages, relay-BP and MLE, where relay-BP itself has short-budget and long-budget variants. The short-budget and long-budget relay-BP variants use different parameters, which correspond to different computational costs. After the short stage returns, the syndrome condition \(H_{\mathrm{det}}\hat e=s\) is checked independently by binary matrix multiplication. If it holds, the sample is accepted immediately. Otherwise the long stage is run with the same syndrome and the original prior, without inheriting messages, posterior, or correction from the short stage. If the long stage also fails the syndrome check, the first stage abstains. 

Samples for which the first stage abstains are passed to the MLE decoder. The term MLE here refers to single-pattern MLE, not to the class with the largest total probability. Degeneracy makes the two objectives differ in general~\cite{landahl2011color,bravyi2014mld}. In the independent detector error model used for decoding, an error pattern \(e\in\{0,1\}^J\) has probability
\begin{equation}
P(e)=\prod_{j=1}^{J}p_j^{e_j}(1-p_j)^{1-e_j}.
\label{eq:pattern}
\end{equation}
For a given syndrome \(s\), maximizing \(P(e)\) is equivalent to minimizing the negative log prior. Let \(d_i\) be the number of nonzeros in row \(i\) of \(H_{\mathrm{det}}\), and let the integer variable \(z_i\) represent the parity slack. The MLE stage solves
\begin{equation}
\min_{e,z}\ \sum_{j=1}^{J}\lambda_j e_j,
\qquad
\lambda_j=\log\frac{1-p_j}{p_j},
\label{eq:milp}
\end{equation}
subject to
\begin{equation}
\begin{aligned}
   H_{\mathrm{det}}e-2z & =s,\qquad
e_j\in\{0,1\},\quad \\
z_i & \in\{0,\ldots,\lfloor d_i/2\rfloor\}. 
\end{aligned}
\label{eq:milp-constraints}
\end{equation}
The equality holds over the integers, and uses the original prior probabilities, not the Relay-BP posterior.

Statistics are collected separately for each code with $X$ or $Z$ memory basis. We sample at the physical error rate \(p_{\mathrm{phy}}=0.001\), keeping the numbers of \(X\)- and \(Z\)-basis samples equal. We report the normalized logical error rate per logical qubit per QEC round as
\begin{equation}
P_{\mathrm{L}}=1-\left(1-P_{\mathrm{fail}}\right)^{1/(n_r k)}.
\label{eq:rates}
\end{equation}
Here, \(P_{\mathrm{fail}}\) is the probability that at least one logical observable in the code block is decoded incorrectly over the \(n_r\)-round memory experiment. For \(P_{\mathrm{fail}}\ll1\), this definition reduces to \(P_L\simeq P_{\mathrm{fail}}/(n_rk)\). The per-round normalization is motivated by the approximately linear accumulation of rare memory failures when the logical failure rate is approximately stationary.

\begin{figure}[!t]
\centering
\includegraphics[width=\columnwidth]{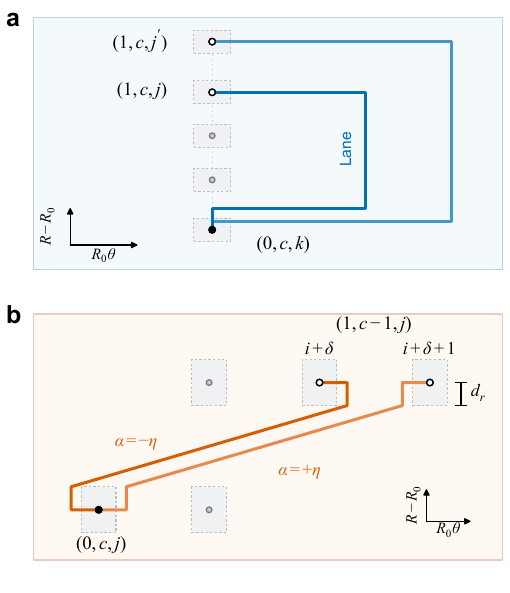}
\caption{\textbf{Local paths for radial and circumferential connections.}
The blue and orange panels illustrate the two connection families in Fig.~\ref{fig:construction}. Their colors distinguish connection types, not routing faces.
\textbf{a,} Two radial paths at the same station reach different channels, using separate angular lanes to pass intermediate ports.
\textbf{b,} Two circumferential paths with neighboring shifts $\delta$ and $\delta+1$ share a face. In the illustrated case, opposite endpoint offsets $\alpha=\mp\eta$ with $\eta=1/4$ make their middle segments parallel. The optimized layouts also allow smaller local offsets. The radial leads clear the port region of half-width $d_r$. Filled and open circles denote $r=0$ and $r=1$, and rail labels omit the station index $i$. Dashed gray rectangles indicate schematic local port regions and unused ports are faded. The axes show unfolded coordinates, scaled independently in the two directions.}
\label{fig:routing}
\end{figure}

\vspace{.2cm}
\noindent\textbf{Geometric layout for the multi-tier architecture} \\
\vspace{.2cm}
The pairing rules determine which physical qubits must interact, while leaving freedom in the shapes of the connecting paths and their assignment to routing faces. We use this freedom to reduce wire lengths and vertical interconnects without changing the code or the syndrome-extraction circuit. The annular geometry provides local paths for radial and circumferential connections. Their dimensions are adapted to the connections required by each code, and their compatibility determines how they share the multi-tier structure.

Each physical module contains the positions $(0,c,i,j)$ and $(1,c-1,i,j)$, with $c$ taken modulo $M$. Using the $r=0$ value of $c$ to label each module, their order from the inner to the outer boundary of the annulus is
\begin{equation}
0,\ M-1,\ 1,\ M-2,\ \ldots,\ M/2-1,\ M/2.
\end{equation}
For $M=12$, this gives $(0,11,1,10,2,9,3,8,4,7,5,6)$. Within each module, the channels are arranged outwards in increasing $j$. For $c<M/2$, the position $(1,c-1,i,j)$ lies inside $(0,c,i,j)$. For $c\ge M/2$, their radial order is reversed. This arrangement places at most one other module between modules $c$ and $c+1$, including $c=M-1$ and $c+1=0$, limiting the radial span of the even-step connections.

An even-step edge connects $(0,c,i,k)$ and $(1,c,i,j)$ at the same station $\theta_i=2\pi i/p$, as illustrated in \hyperref[fig:routing]{Fig.~\ref*{fig:routing}(a)}. Starting from the $r=0$ endpoint, the path makes a short radial displacement, follows an arc to a nearby angular lane, crosses radially to the target radius, and returns along an arc to the target port. The channel pair $(k,j)$ determines the radial displacement and angular lane. The sign of the angular excursion alternates with $k$, separating paths from neighboring channels before they cross the intervening rails. The panel uses unfolded coordinates $(R_0\theta,R-R_0)$, where $R$ is the physical radius and $R_0$ is the reference radius. Physical lengths and distances are evaluated on the corresponding continuous curves in the annulus.

An odd-step edge connects $(0,c,i,j)$ and $(1,c-1,i+\delta,j)$ within one physical module, with angular separation $\delta\beta$, where $\beta=2\pi/p$. The path begins with an angular offset $\alpha\beta$, moves radially out of the port region, follows a linear segment in the unfolded coordinates, moves radially to the target radius, and follows an arc into the target port. The angular offsets relative to the source and target are $+\alpha\beta$ and $-\alpha\beta$. Hence the middle segment spans $(\delta-2\alpha)\beta$. Adjacent shifts can use opposite offsets to share a face. In the example in \hyperref[fig:routing]{Fig.~\ref*{fig:routing}(b)}, the smaller shift uses $\alpha=-\eta$ and the larger uses $\alpha=+\eta$. Taking $\eta=1/4$ makes the middle segments parallel. In the final layouts, shifts may instead be routed individually or paired with smaller offsets, with separation checked for the resulting complete paths.

The allowed shift ranges constrain the code search, but each module and channel uses only a subset of those shifts. We choose the local radial spacing and endpoint offsets from these actual connections. This avoids reserving space for absent connections and shortens the detours needed to separate nearby paths. The annular ordering and all required endpoint pairs in the circuit are preserved. The physical edge list includes the SWAP-only connections associated with deleted checks.

Face assignment uses the compatibility of the complete paths, including their port leads. We assign all $p$ rotated copies of a connection to the same face and use a conflict graph to identify paths that cannot share a face. A bounded heuristic search reduces the hardware-complexity score while respecting these conflicts and a maximum of four connections per qubit on each face. Radial and circumferential paths may share a face whenever their separation permits it. Every path remains on its assigned face throughout its in-plane route. Thus, the code structure restricts the form of the local routing paths, while the optimization determines how efficiently the available faces are used.

The tier ordering then balances in-plane length and vertical access. For each fixed face assignment, we enumerate the choice of the qubit-tier face and the pairing of the remaining faces into higher tiers. Within each pair, the face with more connections is placed on the lower surface, and pairs with more connections are placed closer to the qubit tier. These choices minimize the vertical-interconnect contributions for that pairing. We select the stacking with the lowest HAL score. The three layouts use 8, 8, and 11 occupied faces, corresponding to 5, 5, and 6 tiers. Note that this is an exhaustive choice of stacking for the selected faces, rather than a proof of global optimality over all geometries and face assignments.

Finally, we independently check the complete continuous paths and all their periodic copies. Paths on the same face may meet only within the fixed regions of their common endpoints. Outside those regions, paths on the same face must remain more than $0.1a_0$ apart, and each complete path must remain more than $0.05a_0$ from every nonendpoint device region, where $a_0$ is the minimum center-to-center spacing of the physical qubits. The resulting layout retains the prescribed annular structure and local routing paths, with optimization confined to their geometric parameters, face assignments, and stacking. The geometric bounds, HAL aggregation, and further layout data are given in the Supplementary Information~\cite{SupplementaryInformation} and \cite{ArmillaryCodeRepository}.
\end{document}

\title{Supplementary Information for ``Rate-1/2 Quantum LDPC Codes for Multi-tier Superconducting Hardware''}
\maketitle

\section{Details of the three representative Armillary codes}
\label{sec:representative-code-details}

Each of the three representative Armillary codes uses $M=12$ data groups and $M=12$ syndrome groups, with $P=qp$ qubits per group. Here $q$ is the number of channels and $p$ is the number of angular stations. The code contains $n=12qp$ data qubits, and the syndrome-extraction circuit uses $2n$ physical positions. Table~\ref{tab:representative-codes} lists the code parameters and retained syndrome groups. The ordered lists $\mathcal X$ and $\mathcal Z$ specify the block-row order of the retained $X$- and $Z$-type syndrome groups. Their underlying sets are denoted by $X$ and $Z$ in the main text. The remaining syndrome groups are used exclusively for routing. They undergo SWAP operations without contributing checks.

\begin{table}[!htbp]
\centering
\small
\setlength{\tabcolsep}{8pt}
\renewcommand{\arraystretch}{1.15}
\begin{tabular*}{0.90\textwidth}{@{\extracolsep{\fill}}cccc@{}}
\toprule
Code $[\![n,k,d]\!]$ & $(q,p)$ & Retained $X$ groups: $\mathcal X$ & Retained $Z$ groups: $\mathcal Z$ \\
\midrule
$[\![3636,1822,12]\!]$ & $(3,101)$ & $(4,0,8)$ & $(6,2,10)$ \\
$[\![4932,2470,14]\!]$ & $(3,137)$ & $(8,0,4)$ & $(2,6,10)$ \\
$[\![8688,4348,16]\!]$ & $(4,181)$ & $(8,4,0)$ & $(10,6,2)$ \\
\bottomrule
\end{tabular*}
\caption{\textbf{Representative Armillary codes and retained syndrome groups.}
The order within $\mathcal X$ and $\mathcal Z$ fixes the block-row order used to reconstruct $H_X$ and $H_Z$.}
\label{tab:representative-codes}
\end{table}

For fixed $q$ and $p$, the discrete search variables are 72 channel permutations $\pi_{h,a}$ and either 72 or 288 integer shifts. In the $q=3$ constructions, all channels share one shift $\delta_{h,a,j}=\delta_{h,a}$ for each $(h,a)$. In the $q=4$ construction, the shifts depend on the current channel $j$. The shifts act modulo $p$ and their signed representatives are restricted to $[-4,4]$ for $q=3$ and $[-6,7]$ for $q=4$ by the prescribed routing library and connection-length constraints. This parameterization replaces a direct search over binary check matrices with thousands of columns by a search over a fixed set of permutations and bounded integers. The parameters are constrained by CSS commutation, $\Omega=0$, and ordered cancellation, $C=0$, with additional screening for four-cycles and low-weight nontrivial logical operators. Each retained $X$- or $Z$-type Tanner graph has girth at least six.

\paragraph{Routing rules and indexing.}
All indices are zero-based. As in the main text, data and syndrome qubits are identified by their functional roles, while physical positions are labeled by $(r,c,i,j)$, where $r\in\{0,1\}$, $c\in\mathbb Z_{12}$, $i\in\mathbb Z_p$, and $j\in\{0,\ldots,q-1\}$. The rail index $r$ records the initial role ($r=0$ for data and $r=1$ for syndrome); $c$, $i$, and $j$ are the group, station, and channel indices, respectively. A data qubit with initial label $(b,i,j)$ starts at $(0,-b,i,j)$, and a syndrome qubit with initial label $(a,i,j)$ starts at $(1,-a,i,j)$. For $h=0,\ldots,5$, the matching at even step $2h$ is
\begin{equation}
(1,c,i,j)\longleftrightarrow
(0,c,i,\pi_{h,a}(j)),
\label{eq:representative-even-matching}
\end{equation}
and the matching at odd step $2h+1$ is
\begin{equation}
(0,c,i,j)\longleftrightarrow
(1,c-1,i+\delta_{h,a,j},j).
\label{eq:representative-odd-matching}
\end{equation}
The group and station indices are reduced modulo 12 and $p$, respectively. In both equations, $a$ denotes the initial group of the syndrome qubit currently at the first endpoint, whereas $j$ denotes its current channel. Thus, the channel selecting an odd-step shift already includes the preceding even-step permutation.

Each matching is determined from the positions at the beginning of the step. On a retained $X$-type check edge, a CX acts from the syndrome qubit to the data qubit; on a retained $Z$-type check edge, the direction is reversed. Each CX is followed by a SWAP, implementing $\mathrm{CXSWAP}=\mathrm{SWAP}\circ\mathrm{CX}$, while unretained syndrome groups undergo SWAP only. All disjoint pairs are updated simultaneously, and the data and syndrome roles are tracked as their states move between physical positions.

Tables~\ref{tab:parameters-3636}, \ref{tab:parameters-4932}, and~\ref{tab:parameters-8688} give all discrete parameters. A permutation is written as its image list $\pi(0)\pi(1)\cdots\pi(q-1)$. For example, \perm{201} denotes $0\mapsto2$, $1\mapsto0$, and $2\mapsto1$. Table rows are indexed by the initial syndrome group $a$ and columns by $h$, corresponding to steps $2h$ and $2h+1$. For $q=4$, each shift entry is the ordered tuple $(\delta_{h,a,0},\delta_{h,a,1},\delta_{h,a,2},\delta_{h,a,3})$.

\paragraph{Check-matrix reconstruction and validation.}
Starting from the initial positions, apply the 12 matchings and record the initial data label encountered by each syndrome qubit before each SWAP. At step $t$, syndrome group $a$ encounters data group $b=a+t\pmod{12}$, so each syndrome qubit visits one data qubit from each of the 12 initial groups. The data-column index $v$ is
\begin{equation}
v=bP+qi+j,\qquad P=qp,
\label{eq:representative-data-index}
\end{equation}
where $(b,i,j)$ is the initial data label, rather than its current physical position. The 12 visited indices define a binary row. Retaining the block rows in the stated order of $\mathcal X$ and $\mathcal Z$, and ordering members within each block by $qi+j$, produces $H_X$ and $H_Z$.

Each check matrix has dimensions $3qp\times12qp$, row weight 12, and column weight three. Its rank over $\mathbb F_2$ is $3qp-2$ for each of the three instances. Consequently,
\begin{equation}
k=n-\operatorname{rank}H_X-\operatorname{rank}H_Z=6qp+4,
\qquad
R=\frac{k}{n}=\frac12+\frac{1}{3qp}.
\end{equation}
The respective matrix dimensions are $909\times3636$, $1233\times4932$, and $2172\times8688$, with rank 907, 1231, and 2170 on each side. The excess over the design rate $1/2$ comes from two linear dependencies in each check matrix. In addition, all code distances reported in the main text have been verified exactly.

The reconstruction checks the complete matchings, row and column weights, the girth of each $X$- and $Z$-type Tanner graph, ranks, CSS commutation, and ordered  cancellation. Writing $H_X(t)$ and $H_Z(t)$ for the step-resolved interaction matrices in the initial role labeling, the latter two conditions are
\begin{equation}
\Omega=H_XH_Z^{\mathsf T}=0,
\qquad
C=\sum_{0\leq t<u<12}H_X(t)H_Z(u)^{\mathsf T}=0,
\end{equation}
with arithmetic over $\mathbb F_2$. The reverse round uses the forward round's matchings in reverse order while retaining the CX-before-SWAP convention and assigning CX directions according to the transported roles. The forward and reverse rounds extract the same checks and together restore all roles to their initial positions. Saved logical witnesses are checked for zero syndrome and exclusion from the corresponding stabilizer row space.

\begin{table}[!htbp]
\centering
\small
\setlength{\tabcolsep}{2.5pt}
\renewcommand{\arraystretch}{1.12}
\begin{minipage}[t]{0.48\textwidth}
\centering
\textbf{(a) Channel permutations $\pi_{h,a}$}
\par\smallskip
\begin{tabular*}{\linewidth}{@{\extracolsep{\fill}}c*{6}{c}@{}}
\toprule
$a$ & $h=0$ & $h=1$ & $h=2$ & $h=3$ & $h=4$ & $h=5$ \\
\midrule
0 & \perm{021} & \perm{021} & \perm{021} & \perm{210} & \perm{012} & \perm{120} \\
1 & \perm{201} & \perm{210} & \perm{210} & \perm{210} & \perm{120} & \perm{120} \\
2 & \perm{201} & \perm{012} & \perm{210} & \perm{021} & \perm{102} & \perm{012} \\
3 & \perm{120} & \perm{120} & \perm{012} & \perm{120} & \perm{210} & \perm{021} \\
\addlinespace[2pt]
4 & \perm{102} & \perm{021} & \perm{102} & \perm{210} & \perm{012} & \perm{102} \\
5 & \perm{012} & \perm{120} & \perm{210} & \perm{102} & \perm{021} & \perm{120} \\
6 & \perm{120} & \perm{012} & \perm{012} & \perm{201} & \perm{102} & \perm{012} \\
7 & \perm{210} & \perm{021} & \perm{201} & \perm{201} & \perm{012} & \perm{012} \\
\addlinespace[2pt]
8 & \perm{102} & \perm{021} & \perm{201} & \perm{021} & \perm{120} & \perm{012} \\
9 & \perm{012} & \perm{102} & \perm{021} & \perm{201} & \perm{012} & \perm{012} \\
10 & \perm{012} & \perm{012} & \perm{210} & \perm{021} & \perm{012} & \perm{120} \\
11 & \perm{021} & \perm{120} & \perm{021} & \perm{012} & \perm{201} & \perm{021} \\
\bottomrule
\end{tabular*}
\end{minipage}
\hfill
\begin{minipage}[t]{0.48\textwidth}
\centering
\textbf{(b) Shared station shifts $\delta_{h,a}$}
\par\smallskip
\begin{tabular*}{\linewidth}{@{\extracolsep{\fill}}c*{6}{r}@{}}
\toprule
$a$ & $h=0$ & $h=1$ & $h=2$ & $h=3$ & $h=4$ & $h=5$ \\
\midrule
0 & $1$ & $1$ & $2$ & $1$ & $3$ & $-1$ \\
1 & $0$ & $3$ & $1$ & $4$ & $-2$ & $-2$ \\
2 & $-1$ & $0$ & $4$ & $2$ & $-1$ & $2$ \\
3 & $4$ & $0$ & $-1$ & $-2$ & $-2$ & $0$ \\
\addlinespace[2pt]
4 & $-1$ & $2$ & $2$ & $0$ & $0$ & $-1$ \\
5 & $-2$ & $-2$ & $0$ & $-1$ & $0$ & $3$ \\
6 & $1$ & $0$ & $0$ & $4$ & $0$ & $1$ \\
7 & $2$ & $1$ & $4$ & $1$ & $3$ & $-1$ \\
\addlinespace[2pt]
8 & $2$ & $1$ & $3$ & $-1$ & $3$ & $-1$ \\
9 & $2$ & $4$ & $-1$ & $-1$ & $3$ & $-2$ \\
10 & $-1$ & $-2$ & $3$ & $4$ & $-2$ & $0$ \\
11 & $2$ & $2$ & $1$ & $2$ & $4$ & $-2$ \\
\bottomrule
\end{tabular*}
\end{minipage}
\caption{\textbf{Routing parameters for the $[\![3636,1822,12]\!]$ Armillary code.} Here $q=3$ and $p=101$. Panel (a) lists the 72 even-step channel permutations $\pi_{h,a}$; panel (b) lists the 72 odd-step station shifts $\delta_{h,a}$ shared by all channels. Row $a$ labels the initial syndrome group, and column $h$ corresponds to steps $2h$ and $2h+1$.}
\label{tab:parameters-3636}
\end{table}

\begin{table}[!htbp]
\centering
\small
\setlength{\tabcolsep}{2.5pt}
\renewcommand{\arraystretch}{1.12}
\begin{minipage}[t]{0.48\textwidth}
\centering
\textbf{(a) Channel permutations $\pi_{h,a}$}
\par\smallskip
\begin{tabular*}{\linewidth}{@{\extracolsep{\fill}}c*{6}{c}@{}}
\toprule
$a$ & $h=0$ & $h=1$ & $h=2$ & $h=3$ & $h=4$ & $h=5$ \\
\midrule
0 & \perm{120} & \perm{201} & \perm{102} & \perm{120} & \perm{012} & \perm{021} \\
1 & \perm{102} & \perm{102} & \perm{012} & \perm{120} & \perm{021} & \perm{021} \\
2 & \perm{102} & \perm{012} & \perm{120} & \perm{021} & \perm{021} & \perm{012} \\
3 & \perm{021} & \perm{120} & \perm{120} & \perm{120} & \perm{201} & \perm{210} \\
\addlinespace[2pt]
4 & \perm{012} & \perm{201} & \perm{120} & \perm{012} & \perm{102} & \perm{021} \\
5 & \perm{210} & \perm{120} & \perm{201} & \perm{021} & \perm{021} & \perm{120} \\
6 & \perm{120} & \perm{201} & \perm{021} & \perm{021} & \perm{120} & \perm{012} \\
7 & \perm{201} & \perm{201} & \perm{201} & \perm{021} & \perm{102} & \perm{012} \\
\addlinespace[2pt]
8 & \perm{102} & \perm{102} & \perm{021} & \perm{012} & \perm{021} & \perm{021} \\
9 & \perm{102} & \perm{201} & \perm{102} & \perm{201} & \perm{021} & \perm{012} \\
10 & \perm{021} & \perm{201} & \perm{201} & \perm{201} & \perm{210} & \perm{102} \\
11 & \perm{210} & \perm{021} & \perm{102} & \perm{102} & \perm{102} & \perm{102} \\
\bottomrule
\end{tabular*}
\end{minipage}
\hfill
\begin{minipage}[t]{0.48\textwidth}
\centering
\textbf{(b) Shared station shifts $\delta_{h,a}$}
\par\smallskip
\begin{tabular*}{\linewidth}{@{\extracolsep{\fill}}c*{6}{r}@{}}
\toprule
$a$ & $h=0$ & $h=1$ & $h=2$ & $h=3$ & $h=4$ & $h=5$ \\
\midrule
0 & $3$ & $-1$ & $-4$ & $-4$ & $1$ & $3$ \\
1 & $4$ & $-3$ & $-4$ & $2$ & $-4$ & $2$ \\
2 & $-2$ & $-3$ & $-1$ & $-3$ & $4$ & $0$ \\
3 & $-1$ & $-1$ & $-3$ & $4$ & $-2$ & $-2$ \\
\addlinespace[2pt]
4 & $1$ & $4$ & $-3$ & $-4$ & $-2$ & $3$ \\
5 & $-3$ & $-4$ & $-4$ & $-3$ & $-2$ & $0$ \\
6 & $2$ & $-1$ & $-4$ & $-3$ & $1$ & $-3$ \\
7 & $-3$ & $2$ & $-3$ & $-4$ & $-2$ & $0$ \\
\addlinespace[2pt]
8 & $3$ & $-3$ & $-4$ & $-4$ & $1$ & $0$ \\
9 & $-2$ & $0$ & $-4$ & $-3$ & $1$ & $-3$ \\
10 & $1$ & $-3$ & $-4$ & $-1$ & $-4$ & $1$ \\
11 & $-4$ & $1$ & $-1$ & $2$ & $3$ & $-3$ \\
\bottomrule
\end{tabular*}
\end{minipage}
\caption{\textbf{Routing parameters for the $[\![4932,2470,14]\!]$ Armillary code.} Here $q=3$ and $p=137$. Panel (a) lists the 72 even-step channel permutations $\pi_{h,a}$. Panel (b) lists the 72 odd-step station shifts $\delta_{h,a}$ shared by all channels. Row $a$ labels the initial syndrome group, and column $h$ corresponds to steps $2h$ and $2h+1$.}
\label{tab:parameters-4932}
\end{table}

\begin{table}[!htbp]
\centering
\small
\setlength{\tabcolsep}{2.5pt}
\renewcommand{\arraystretch}{1.12}
\begin{minipage}[t]{0.70\textwidth}
\centering
\textbf{(a) Channel permutations $\pi_{h,a}$}
\par\smallskip
\begin{tabular*}{\linewidth}{@{\extracolsep{\fill}}c*{6}{c}@{}}
\toprule
$a$ & $h=0$ & $h=1$ & $h=2$ & $h=3$ & $h=4$ & $h=5$ \\
\midrule
0 & \perm{0123} & \perm{0123} & \perm{3012} & \perm{0123} & \perm{2301} & \perm{1320} \\
1 & \perm{0123} & \perm{2130} & \perm{0123} & \perm{2301} & \perm{0123} & \perm{1032} \\
2 & \perm{0123} & \perm{0123} & \perm{1230} & \perm{0123} & \perm{2310} & \perm{3201} \\
3 & \perm{3102} & \perm{1230} & \perm{3021} & \perm{3012} & \perm{3201} & \perm{3012} \\
\addlinespace[2pt]
4 & \perm{1230} & \perm{0123} & \perm{3012} & \perm{1230} & \perm{3012} & \perm{2130} \\
5 & \perm{0123} & \perm{1320} & \perm{0123} & \perm{0123} & \perm{3012} & \perm{3201} \\
6 & \perm{0123} & \perm{0123} & \perm{0123} & \perm{3012} & \perm{0213} & \perm{1320} \\
7 & \perm{3012} & \perm{0123} & \perm{2310} & \perm{1230} & \perm{1320} & \perm{3021} \\
\addlinespace[2pt]
8 & \perm{0123} & \perm{3012} & \perm{1230} & \perm{0123} & \perm{3021} & \perm{0132} \\
9 & \perm{0123} & \perm{3201} & \perm{0123} & \perm{3021} & \perm{1320} & \perm{0231} \\
10 & \perm{0123} & \perm{0123} & \perm{0213} & \perm{1320} & \perm{2301} & \perm{0123} \\
11 & \perm{0123} & \perm{0123} & \perm{0213} & \perm{3012} & \perm{1023} & \perm{2301} \\
\bottomrule
\end{tabular*}
\end{minipage}
\par\medskip
\begin{minipage}[t]{0.49\textwidth}
\centering
\textbf{(b) Station shifts, $h=0,1,2$}
\par\smallskip
\begin{tabular*}{\linewidth}{@{\extracolsep{\fill}}c*{3}{c}@{}}
\toprule
$a$ & $h=0$ & $h=1$ & $h=2$ \\
\midrule
0 & \shifttuple{3}{-1}{0}{-2} & \shifttuple{7}{5}{-1}{7} & \shifttuple{-3}{2}{1}{-4} \\
1 & \shifttuple{5}{2}{7}{5} & \shifttuple{-5}{0}{-3}{-6} & \shifttuple{5}{4}{3}{5} \\
2 & \shifttuple{2}{7}{-3}{7} & \shifttuple{7}{6}{7}{7} & \shifttuple{-3}{-3}{-1}{-1} \\
3 & \shifttuple{3}{2}{4}{3} & \shifttuple{-4}{-6}{0}{-2} & \shifttuple{7}{6}{5}{0} \\
\addlinespace[2pt]
4 & \shifttuple{2}{1}{3}{3} & \shifttuple{7}{7}{3}{1} & \shifttuple{0}{-3}{2}{1} \\
5 & \shifttuple{5}{7}{0}{7} & \shifttuple{2}{-1}{5}{4} & \shifttuple{-3}{1}{2}{2} \\
6 & \shifttuple{6}{7}{7}{7} & \shifttuple{3}{3}{3}{-1} & \shifttuple{2}{4}{1}{2} \\
7 & \shifttuple{7}{7}{7}{2} & \shifttuple{6}{-2}{3}{2} & \shifttuple{-3}{1}{2}{1} \\
\addlinespace[2pt]
8 & \shifttuple{7}{-4}{4}{4} & \shifttuple{4}{0}{2}{-1} & \shifttuple{-4}{-5}{-4}{-2} \\
9 & \shifttuple{0}{-6}{-6}{0} & \shifttuple{6}{3}{4}{6} & \shifttuple{-2}{2}{5}{-4} \\
10 & \shifttuple{7}{3}{7}{1} & \shifttuple{-2}{2}{5}{-4} & \shifttuple{3}{3}{5}{5} \\
11 & \shifttuple{-4}{1}{4}{-5} & \shifttuple{6}{2}{4}{6} & \shifttuple{2}{1}{-3}{-6} \\
\bottomrule
\end{tabular*}
\end{minipage}
\hfill
\begin{minipage}[t]{0.49\textwidth}
\centering
\textbf{(c) Station shifts, $h=3,4,5$}
\par\smallskip
\begin{tabular*}{\linewidth}{@{\extracolsep{\fill}}c*{3}{c}@{}}
\toprule
$a$ & $h=3$ & $h=4$ & $h=5$ \\
\midrule
0 & \shifttuple{5}{2}{3}{1} & \shifttuple{5}{7}{7}{7} & \shifttuple{0}{-2}{-4}{2} \\
1 & \shifttuple{-3}{-3}{-1}{-1} & \shifttuple{6}{1}{7}{-1} & \shifttuple{0}{0}{0}{0} \\
2 & \shifttuple{6}{4}{4}{0} & \shifttuple{2}{-2}{2}{2} & \shifttuple{2}{6}{4}{7} \\
3 & \shifttuple{7}{-4}{-3}{4} & \shifttuple{-2}{2}{0}{4} & \shifttuple{0}{0}{0}{0} \\
\addlinespace[2pt]
4 & \shifttuple{-3}{1}{2}{2} & \shifttuple{6}{7}{5}{7} & \shifttuple{7}{4}{5}{7} \\
5 & \shifttuple{4}{5}{2}{4} & \shifttuple{7}{7}{3}{5} & \shifttuple{0}{0}{0}{0} \\
6 & \shifttuple{-2}{2}{2}{3} & \shifttuple{-6}{-2}{-5}{-6} & \shifttuple{3}{3}{3}{4} \\
7 & \shifttuple{-3}{-3}{-6}{-3} & \shifttuple{7}{7}{7}{7} & \shifttuple{0}{0}{0}{0} \\
\addlinespace[2pt]
8 & \shifttuple{0}{4}{1}{2} & \shifttuple{3}{3}{5}{5} & \shifttuple{5}{6}{7}{4} \\
9 & \shifttuple{5}{7}{7}{7} & \shifttuple{2}{6}{6}{7} & \shifttuple{0}{0}{0}{0} \\
10 & \shifttuple{2}{1}{-3}{-6} & \shifttuple{3}{7}{7}{5} & \shifttuple{0}{-2}{-6}{-4} \\
11 & \shifttuple{5}{7}{6}{2} & \shifttuple{1}{-1}{-5}{-2} & \shifttuple{0}{0}{0}{0} \\
\bottomrule
\end{tabular*}
\end{minipage}
\caption{\textbf{Routing parameters for the $[\![8688,4348,16]\!]$ Armillary code.} Here $q=4$ and $p=181$. Panel (a) lists the 72 even-step channel permutations $\pi_{h,a}$. Panels (b) and (c) list the 288 odd-step station shifts as ordered tuples $(\delta_{h,a,0},\delta_{h,a,1},\delta_{h,a,2},\delta_{h,a,3})$ for the current channels. Row $a$ labels the initial syndrome group, and column $h$ corresponds to steps $2h$ and $2h+1$.}
\label{tab:parameters-8688}
\end{table}

\clearpage
\section{Optimized hardware layouts}
\label{sec:optimized-hardware-layouts}

The layouts preserve the code and syndrome-extraction circuit. Physical connections are the distinct pairs used in the 12 steps, including SWAP-only connections associated with unretained syndrome groups. Repeated uses require only one route. Path shapes and routing faces are optimized while retaining the folded module order and angular translation symmetry. All $p$ station copies of a connection use the same local path and face.

Lengths are expressed in units of the minimum distance $a_0$ between any two of the $2n$ physical qubit positions. Each position has a fixed device region, defined in the unfolded coordinates $u=R_0\theta$ and $R$ by $|u-u_{\mathrm c}|\leq d_u$ and $|R-R_{\mathrm c}|\leq d_r$, where $(u_{\mathrm c},R_{\mathrm c})$ is the center of that device. The corresponding region in the annulus is obtained by the polar map. The half-widths in Table~\ref{tab:optimized-layouts} are held fixed for each code. The parameter $d_u$ is measured at the reference radius, so it is not a constant physical arc half-width at all radii. Outside their common endpoint regions, paths on the same face must be separated by more than $0.1a_0$. Every complete path must also remain more than $0.05a_0$ from all device regions except those of its own endpoints.

\begin{table}[htb]
\centering
\small
\setlength{\tabcolsep}{7pt}
\renewcommand{\arraystretch}{1.12}
\begin{tabular}{@{}lccc@{}}
\toprule
 & $[\![3636,1822,12]\!]$ & $[\![4932,2470,14]\!]$ & $[\![8688,4348,16]\!]$ \\
\midrule
Occupied routing faces & 8 & 8 & 11 \\
Physical connections & 24,038 & 32,195 & 66,065 \\
Mean connection length$/a_0$ & 9.3237 & 10.3428 & 18.8947 \\
Maximum connection length$/a_0$ & 18.8002 & 19.0539 & 48.4811 \\
Wire separation lower bound$/a_0$ & 0.100708 & 0.100539 & 0.100097 \\
Device clearance lower bound$/a_0$ & 0.097228 & 0.099916 & 0.065302 \\
Device-region half-widths $(d_r,d_u)/a_0$ & $(0.152,0.405)$ & $(0.152,0.3675)$ & $(0.35,1.005)$ \\
\bottomrule
\end{tabular}
\caption{\textbf{Geometry of the optimized hardware layouts.} Lengths include the complete continuous path and its endpoint leads. The mean gives equal weight to every physical connection, unlike the tier-averaged Length in the main text. The wire separation and device clearance bounds are rounded down. Device clearance refers to the distance from a path to a nonendpoint device region.}
\label{tab:optimized-layouts}
\end{table}

The paths retain the radial and circumferential forms described in the Methods. Radial paths use opposite angular directions for even and odd channel indices at the $r=0$ end. Circumferential paths use the shifts present locally, routed individually or in adjacent pairs, with radial spacing adjusted accordingly. Individual shifts use zero endpoint angular offset. Paired paths use offsets of magnitude $1/4$ of the station separation in the $q=4$ layout. The $q=3$ layouts also use smaller offsets to shorten the leads. Their middle segments need not be parallel, and separation is checked for the complete curves.

Face assignments respect the separation requirements and allow at most four connections at any endpoint on one face. Radial and circumferential paths may share a face, with no intermediate face switches. The eight, eight, and eleven occupied faces fit within five, five, and six tiers, respectively. For fixed face assignments, we enumerate the qubit-tier face and all pairings of the remaining faces, giving 840 stackings for each eight-face layout and 10,395 for the eleven-face layout. Within each higher tier, the more populated face occupies the lower surface. More populated tiers lie closer to the qubit tier. These choices minimize the Bonds and TSV contributions for each pairing. The reported stacking minimizes the HAL score for the chosen faces. Path and face optimization is heuristic and does not establish global optimality.

Reproducible layout results for the three representative codes are provided in \cite{ArmillaryCodeRepository}. We use the HAL aggregation rules~\cite{halpaper} with continuous arc lengths normalized by $a_0$. Length is the unweighted mean of the within-tier mean connection lengths, including the qubit tier. In contrast, the mean in Table~\ref{tab:optimized-layouts} weights all physical connections equally. Thus, redistributing unchanged paths among tiers can change Length without changing the physical mean. The score is
\begin{equation}
C_{\mathrm{hw}}=1+\frac{T-1}{16}
+\frac{\max(0,\mathrm{Length}-1)}{36}
+\frac{\mathrm{TSVs}}{12}+\frac{\mathrm{Bonds}}{16},
\end{equation}
where $T$ is the number of tiers. TSVs averages the TSV count over connections above the qubit tier. Bonds is the maximum, over tiers, of the mean face-switch segment count in the HAL implementation. It is not a physical bump-bond count. Avoiding intermediate face switches does not remove the vertical connections needed to reach a higher tier or its upper surface.

The separation bounds are obtained from complete continuous curves, including the endpoint leads. The checks cover all same-face path pairs and station translations. Only the fixed regions of shared endpoints are excluded from a pairwise wire check. Exact segment distances in the unfolded coordinates are combined with analytic bounds for the polar map and more distant periodic copies. The verification therefore tests continuous separation rather than distances between sampled points. It uses double-precision arithmetic and does not constitute a formal interval-arithmetic proof. The bounds describe centerlines and fixed device regions, without a manufacturing tolerance or a model of shielding, vertical-interconnect footprints, or control and readout wiring.

The check matrices, physical pairings, and gate order are unchanged, so the circuit-level results apply under the stated noise model. They do not account for length-dependent gate errors or constitute an electromagnetic simulation of the optimized layouts.

\section{Parameters of the relay-BP decoding stage}
\label{sec:relay-bp-parameters}

The first stage of the two-stage decoder described in the main text uses relay-BP~\cite{muller2025relay} with short-budget and long-budget runs. The same parameter settings are used for the $X$- and $Z$-basis memory experiments of all codes. The decoder receives the observed detector syndrome $s$, the detector matrix $H_{\mathrm{det}}$, and the error probabilities obtained from the circuit detector error model. Each sample is first processed by the short-budget run. Its estimated error pattern $\hat e$ is accepted if an independent check confirms
\begin{equation}
H_{\mathrm{det}}\hat e=s \pmod 2.
\label{eq:relay-syndrome-consistency}
\end{equation}
Otherwise, the long-budget decoder is run on the same syndrome and original error priors, with a fresh decoder and no messages or posterior information inherited from the short-budget run. The first syndrome-consistent estimate is accepted. If neither run returns such an estimate, the relay-BP stage abstains and the sample is passed to the second-stage most-likely-error (MLE) decoder described in the main text. 

\begin{table}[!htbp]
\centering
\small
\setlength{\tabcolsep}{8pt}
\renewcommand{\arraystretch}{1.18}
\begin{tabular*}{0.76\textwidth}{@{\extracolsep{\fill}}lcc@{}}
\toprule
Parameter & Short-budget run & Long-budget run \\
\midrule
\texttt{pre\_iter} & 200 & 500 \\
\texttt{num\_sets} & 20 & 200 \\
\texttt{set\_max\_iter} & 100 & 200 \\
\texttt{gamma0} & 0.125 & 0.35 \\
\addlinespace[3pt]
\texttt{stop\_nconv} & 2 & 2 \\
\texttt{seed} & 0 & 0 \\
\bottomrule
\end{tabular*}
\caption{\textbf{Parameters of the relay-BP decoding stage.}
The settings are shared by the $X$- and $Z$-basis memory experiments of all codes.
The implementation uses \texttt{RelayDecoderF64} from Relay-BP version 0.2.2;
unlisted parameters retain the defaults of that version.}
\label{tab:relay}
\end{table}

Table~\ref{tab:relay} specifies the iteration budgets and initial memory weights. The parameter \texttt{pre\_iter} limits the BP iterations in the initial ensemble, \texttt{num\_sets} limits the number of subsequent relay ensembles, and \texttt{set\_max\_iter} limits the iterations per relay ensemble. The parameter \texttt{gamma0} sets the uniform memory weight used in the initial ensemble. These budgets are upper limits; the library may terminate earlier according to its stopping criterion.

The implementation uses \texttt{RelayDecoderF64} from Relay-BP version 0.2.2~\cite{relaybpsoftware022}. Both relay-BP variants use \texttt{stop\_nconv=2} and \texttt{seed=0}, with a fresh decoder instantiated for every sample and every invoked run. The value of \texttt{stop\_nconv} overrides the library default of 1. All remaining settings retain the defaults of this version. The parameter settings are fixed before sampling.

\section{Circuit-level decoding results}
\label{sec:circuit-decoding-results}

We evaluate the three representative Armillary codes at $p_{\mathrm{phy}}=10^{-3}$ with four noisy QEC rounds and ideal boundary operations, using the circuit noise model specified in the main text. Each sample is an $X$- or $Z$-basis memory experiment on the full code block. The relay-BP settings are those of Table~\ref{tab:relay}. Samples for which relay-BP abstains are processed by the MLE decoder using the original detector error probabilities. Here MLE names the most-likely-error objective in the main text. 

\subsection{Sample accounting and decoding outcomes}

For each memory basis, let $N$ be the number of completed samples, $C_1$ the number decoded correctly by relay-BP, $W_1$ the number of incorrect relay-BP outputs, and $B$ the number passed to MLE. Let $C_2$ and $W_2$ count the correct and incorrect MLE outputs. Thus,
\begin{equation}
N=C_1+W_1+B,\qquad B=C_2+W_2,\qquad F=W_1+W_2,
\label{eq:supp-decoding-counts}
\end{equation}
where $F$ is the number of final block failures. A sample fails if at least one logical observable differs from its simulated value. Syndrome consistency alone does not imply correct decoding, so an incorrect relay-BP output contributes to $F$ even though it is not passed to MLE.

Table~\ref{tab:decoding-counts} includes completed samples from matched $X$- and $Z$-basis sampling batches, including completed pairs from interrupted runs. Each sample is counted once and scored using its final correction.

\begin{table}[!htbp]
\centering
\small
\setlength{\tabcolsep}{6pt}
\renewcommand{\arraystretch}{1.15}
\begin{tabular}{ccrrrrrr}
\toprule
$d$ & Basis & $N$ & $C_1$ & $W_1$ & $B$ & $C_2$ & $W_2$ \\
\midrule
12 & $X$ & 3,361,646 & 3,361,385 & 13 & 248 & 248 & 0 \\
   & $Z$ & 3,361,646 & 3,361,385 & 9 & 252 & 252 & 0 \\
\addlinespace[3pt]
14 & $X$ & 10,386,959 & 10,386,478 & 3 & 478 & 478 & 0 \\
   & $Z$ & 10,386,959 & 10,386,431 & 6 & 522 & 522 & 0 \\
\addlinespace[3pt]
16 & $X$ & 17,742,045 & 17,740,645 & 0 & 1,400 & 1,400 & 0 \\
   & $Z$ & 17,742,045 & 17,740,002 & 1 & 2,042 & 2,042 & 0 \\
\midrule
\multicolumn{2}{c}{Total} & 62,981,300 & 62,976,326 & 32 & 4,942 & 4,942 & 0 \\
\bottomrule
\end{tabular}
\caption{\textbf{Decoding outcomes at $p_{\mathrm{phy}}=10^{-3}$.} Distances 12, 14, and 16 identify the $[\![3636,1822,12]\!]$, $[\![4932,2470,14]\!]$, and $[\![8688,4348,16]\!]$ codes, respectively. Counts refer to four-round memory samples. The MLE columns report only logical correctness. All 4,942 samples passed to MLE have a scored final correction.}
\label{tab:decoding-counts}
\end{table}

The distance-16 data combine three sampling batches. The archived count of all completed samples is $35{,}484{,}282$. To retain equal sampling of the two memory bases, we exclude 192 completed but unpaired samples, all correctly decoded by relay-BP, giving $N=35{,}484{,}090$ in the main-text decoding figure and the tables here. This exclusion changes neither the failure counts nor the 3,442 samples passed to MLE. A separate set of 192 samples without a completed first-stage result is excluded from both counts and is not treated as successful decoding.

\subsection{Rate normalization and counting uncertainty}

For each code, we combine the equal-sized $X$ and $Z$ memory data and estimate the block failure probability as $ P_{\mathrm{fail}}=F/N$. As in the main text, we report the equivalent rate
\begin{equation}
 P_{\mathrm L}
=1-\left(1-\frac{F}{N}\right)^{1/(4k)}
\simeq \frac{F}{4kN}.
\label{eq:supp-equivalent-rate}
\end{equation}
This is a normalization of the probability that the block fails. It does not assume that the $k$ logical observables actually fail independently.

\begin{table}[!htbp]
\centering
\small
\setlength{\tabcolsep}{7pt}
\renewcommand{\arraystretch}{1.15}
\begin{tabular}{crrcc}
\toprule
$d$ & $N$ & $F$ & $\widehat P_{\mathrm L}$ & Nominal 95\% reference interval \\
\midrule
12 & 6,723,292 & 22 & $4.49\times10^{-10}$ & $[2.81,\,6.80]\times10^{-10}$ \\
14 & 20,773,918 & 9 & $4.38\times10^{-11}$ & $[2.01,\,8.32]\times10^{-11}$ \\
16 & 35,484,090 & 1 & $1.62\times10^{-12}$ & $[0.0410,\,9.03]\times10^{-12}$ \\
\bottomrule
\end{tabular}
\caption{\textbf{Block-normalized logical error rates.} Rates use Eq.~\eqref{eq:supp-equivalent-rate}. The intervals describe nominal binomial counting uncertainty and do not correct for the sampling stopping rules.}
\label{tab:decoding-rates}
\end{table}

The error bars are obtained by applying Eq.~\eqref{eq:supp-equivalent-rate} to the endpoints of the two-sided 95\% Clopper--Pearson interval for $F$ failures in $N$ trials. They provide a fixed-size binomial reference for the pooled memory data. 

\bibliography{ref}